\documentclass[runningheads]{llncs}
\usepackage[utf8]{inputenc}
\usepackage{graphicx}
\usepackage{hyperref}

\def\equationautorefname~#1\null{Eq.~(#1)\null}

\usepackage{tipa}
\usepackage{todonotes}
\presetkeys{todonotes}{color=yellow, noinline, caption={}, size=\scriptsize}{}

\usepackage{xspace}
\newcommand{\xiplot}{$\chi$iplot\xspace}

\begin{document}
\title{\xiplot: web-first visualisation platform for multidimensional data%
\thanks{Supported by the Research Council of Finland (decisions 346376 and 345704) and the Future Makers Funding Programme of Technology Industries of Finland Centennial Foundation and Jane and Aatos Erkko Foundation.}}
\titlerunning{\xiplot}
\author{Akihiro Tanaka\inst{1}\orcidID{0000-0003-2363-9790} \and
Juniper Tyree\inst{1}\orcidID{0000-0002-7923-9609} \and
Anton Bj\"orklund\inst{1}\orcidID{0000-0002-7749-2918} \and
Jarmo M\"akel\"a\inst{1,2}\orcidID{0000-0002-8788-3939} \and
Kai Puolam\"aki\inst{1}\orcidID{0000-0003-1819-1047}}
\authorrunning{A. Tanaka et al.}
\institute{University of Helsinki, Helsinki, Finland \\
\email{firstname.lastname@helsinki.fi}\and
CSC -- IT Center for Science Ltd, Espoo, Finland \\
\email{firstname.lastname@csc.fi}
}
\maketitle              % typeset the header of the contribution
\noindent This is the author’s version of the work.
The definitive Version of Record will be published in the Proceedings of ECML PKDD 2023.
% The Version of Record of this contribution is published in [volume title], and is available online at [DOI of the paper] 

\begin{abstract}

\xiplot is an HTML5-based system for interactive exploration of data and machine learning models.
A key aspect is interaction, not only for the interactive plots but also between plots. 
Even though \xiplot is not restricted to any single application domain, we have developed and tested it with domain experts in quantum chemistry to study molecular interactions and regression models.
\xiplot can be run both locally and online in a web browser (keeping the data local).
The plots and data can also easily be exported and shared.
A modular structure also makes \xiplot optimal for developing machine learning and new interaction methods.

\end{abstract}

\section{Introduction and related work}\label{sec:intro}

This paper introduces \xiplot (\textipa{\textvertline"kaIpl6t\textvertline}), a modular system for interactive exploration of data and pre-trained machine learning models.
\xiplot can be run locally on the user's computer or installation-free in a web browser. Our motivation for writing \xiplot was three-fold.

(i) First, we want a Python-based system to develop and test machine learning and dimensionality reduction methods, such as \cite{bjorklund2023slisemap}, a manifold visualisation method for explainable AI.
For this purpose, we prefer a {\em modular} system that is easy to expand and modify to test new machine learning and visualisation methods and interaction ideas.

(ii) Second, we need a tool to facilitate collaboration with primarily domain experts in quantum chemistry but also other domains. Ideally, we want to avoid forcing our collaborators to install additional software. However, we also do not want to set up and maintain server infrastructure to host a web-accessible service. 

(iii) Third, the system should be practical and usable for the end user, including physicists and chemists, despite being built for quick prototyping and painless implementation. We know no prior system satisfies all of these three requirements.

Many interactive visualisation tools are available; see, e.g., \cite{qin2020Making} for a recent survey and references.
Much of our research collaboration targets quantum chemistry; hence the system must also be capable of visualising, e.g., molecular structures from SMILES strings \cite{weininger1988smiles}. ChemInformatics Model Explorer\cite{humer2022cime} (CIME) is another tool that explores explainable AI in small molecule research.
However, CIME has only four fixed views, and full functionality requires a server. 
Another recent example is XSMILES\cite{heberle2023xsmiles}, where users can examine individual molecules in 2D diagrams and visualise attribution scores for atoms and non-atom tokens.

\section{Usage}\label{sec:usage}

The main idea of \xiplot is to simultaneously show multiple plots and visualisations to compare and contrast diverse information. 
Since \xiplot also targets non-technical end users, intuitive visual selection and configuration of the plots are required.

\xiplot comes with six types of plots out-of-the-box -- scatterplots, histograms, heat maps, bar plots, data tables, and SMILES plots, which render molecules in a stick structure from a SMILES string\cite{weininger1988smiles} -- but more can be added with \xiplot's plugin system. Users can add and remove plots to create a layout that is the most optimal for their specific needs. 
The end users have the capability to generate clusters by running a k-means algorithm or by lasso selection on a scatterplot. Unique colours distinguish the generated clusters. In addition, the end users can generate a 2D embedding through Principal Component Analysis (PCA).

To use \xiplot, the user may install it with \texttt{pip install xiplot}. The \texttt{xiplot} console command is then available to host a local \xiplot server. 
Alternatively, an \emph{installation-free} WebAssembly (WASM)\footnote{WASM is supported in most modern browsers; see \url{https://caniuse.com/wasm}.} version can be used immediately at \url{https://edahelsinki.fi/xiplot}.

\begin{figure}[ht]
    \centering
    \includegraphics[width=\textwidth]{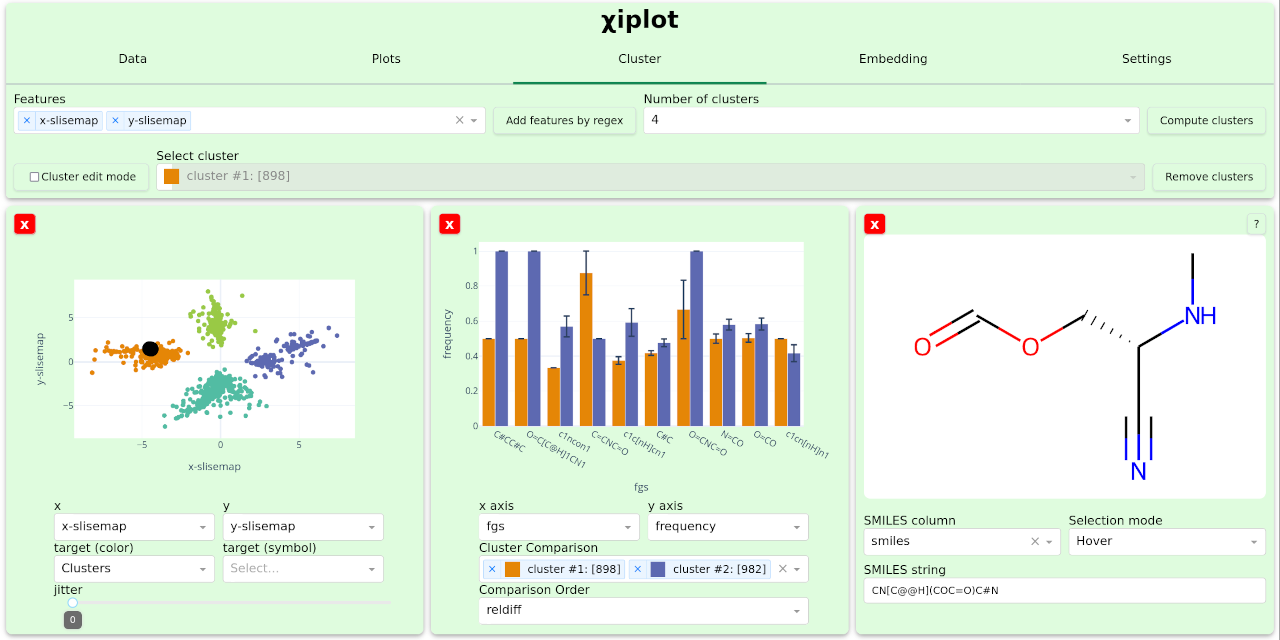}
    \caption{\xiplot interface when studying a regression model on a QM9 dataset.}
    \label{fig:xiplot}
\end{figure}

We demonstrate the main concepts with the QM9 molecular dataset
\cite{ramakrishnan2014Quantum,stuke2019chemical}, a collection of quantum chemical properties calculated for small organic molecules. Our machine-learning task is to estimate some quantum chemical properties from their structural description.
We can use physics simulators with varying fidelity or regression models. 
In this example,
we want to study how the structures in the dataset relate to the estimation task.
We have precomputed a 2D Slisemap\cite{bjorklund2023slisemap} embedding (revealing the structures relevant to a regression model) and attached the embedding to the dataset file we uploaded to \xiplot. 

\autoref{fig:xiplot} shows a view of the \xiplot interface during our exploration.
A chemist can explore the Slisemap embedding in a scatter plot on the left.
There is a notable cluster structure, so we use \xiplot to find the clusters and plot their distribution in the middle.
If we compare the two clusters, we notice that the distributions of the functional groups differ.
For example, we could manually draw an additional cluster in the scatter plot to further study the two subgroups in the rightmost cluster.

The behaviour of a molecule is not only determined by the functional groups but also by how they are structured.
However, finding good summary statistics for structure is much more difficult.
Therefore, we add a visualisation of individual molecules on the right of \autoref{fig:xiplot}.
A chemist can then rapidly inspect multiple molecules inside and between clusters by hovering over the points in the scatter plot; the molecule visualisation is automatically updated.

\section{Description of the system}\label{sec:description}

A key aspect of \xiplot is interactivity, not just for a single plot but also between plots. For example, selecting a data item in one might show you more information about it in another, as described above.
To accomplish this interactivity, the plots of \xiplot are implemented as independent modules, communicating through shared data storage. Furthermore, to support collaboration and sharing, the set of active plots, their configuration, and the data can be saved to and restored from a file. 
Since \xiplot is an interactive system, time-consuming computations (e.g., learning the Slisemap embedding) should be done as part of data preprocessing.

\xiplot is implemented in \texttt{Python} using \texttt{Plotly}\cite{plotly} for the plots and \texttt{Dash} for the interactivity. 
Usually, this would require the users to be able to install \texttt{Python} packages (see \autoref{sec:usage}). However, we also provide a static server-less webpage version of \xiplot that runs both the \texttt{Dash} backend and the \texttt{Plotly} frontend installation-free inside a browser using WebAssembly\cite{webassembly} (WASM).
This also means no data leaves the user's computer in the WASM version.

In detail, the WASM version of \xiplot uses \texttt{Pyodide}\cite{pyodide} to run Python in the browser. The front- and backend communication is intercepted and redirected to the in-WASM server, inspired by the \texttt{WebDash} prototype \cite{webdash}. Crucially, neither the front- nor backend code needs to know that it runs inside a browser. 

As Pyodide does not yet support all Python packages, we use dynamic import detection to enable certain features and fallbacks, such as additional data file formats.
Deploying the WASM version requires bundling all frontend files, \xiplot, and the scripts that bootstrap the web app in the WASM backend, all documented in the \xiplot GitHub repository.

To open up \xiplot to even more use cases, \xiplot has an API for creating plugins for, e.g., new visualisations and machine learning methods.
It uses the ``entry points'' feature of \texttt{Python} to discover installed plugins, which also works in the WASM version.
Due to the modular design with shared data, new plots can automatically interact with old ones.

\section{Conclusions}\label{sec:conclusions}

We have already found \xiplot helpful when collaborating with domain experts since it lets them configure interactive plots without programming or installing anything\footnote{Installation-free version at \url{https://edahelsinki.fi/xiplot}}.
The online version also enables easy results sharing without exposing the data to any third party.
For more technical users \xiplot is easy to maintain end expand due to the modular architecture.
Finally, \xiplot is available under the Open Source MIT license from GitHub\footnote{\url{https://github.com/edahelsinki/xiplot}} (which includes documentation, usage examples, and a demonstration video).

\bibliographystyle{splncs04}
\bibliography{ms}

\end{document}